\documentstyle[12pt]{article}
\begin{document}
\title
{Discrete Symmetries Underlying \\ Some Continuous Ones:
Two Examples \\ From Gravity And Particle Physics }
\author
{By Michael A. Ivanov \\
Chair of Physics, \\
Belarus State University of Informatics and Radioelectronics, \\
6 P. Brovka Street,  BY 220027, Minsk, Republic of Belarus.\\
E-mail: ivanovma@gw.bsuir.unibel.by.}

\maketitle

\begin{abstract}
Two examples, not connected at present, from author's papers 
(Nuovo Cim., 1992, v.105A, p.77 [hep-th/0207210] and 
GRG, 1999, v.31, p.1431 [gr-qc/0207017]) are considered here in which
a physical model has discrete symmetries and additional non-observable
coordinates or parameters. Then it is possible to introduce some apparent
continuous symmetries of the model for an observer which cannot know 
values of these additional quantities.
\end{abstract}

\section[1]{Introduction }
In the standard model of particle physics \cite{1}, we see a very 
complecated primary postulate about the kind of continuous internal
symmetry group. In addition, multiplets in different generations 
should be transformed by the interwoven representations of the
group. It is obvious that such the complecated postulates are not 
good for a fundamental theory. 
\par
In this paper, the author describes two own examples from particle
physics and gravity in which only some kind of discrete symmetry
takes place for the initial model equations but one has a 
possibility to introduce a continuous symmetry on this base.
There is a common feature for the both examples: one must have 
some additional coordinates or parameters to assume the new simple 
demands on solutions with a discrete symmetry obeying transition
to a continuous one.
\par
As the first example, we consider here a model of composite fermions
\cite{2}. The second example is an embedding the general relativity
$4-$space into a flat $12-$space that results in an alternative
model of gravity \cite{3}.

\section[2]{Symmetries of the composite fermions \\ model \cite{2}}

By linearization of an equation for an energy $E$ of a 
two-component system
$$E  = (m_{1}^{2}+\bar{p}_{1}^{2})^{1/2} + (m_{2}^{2}+\bar{p}
 _{2}^{2})^{1/2}$$
(and assuming after it that constituents' masses $m_{1}= m_{2}=0$;
$\bar{p}_{1}$ and $\bar{p}_{2}$ are their momentums), one can
get the system of linear quantum equations (13)-(16) from \cite{2}
for a wave function $\psi(x_{1},x_{2}),$ where $x_{1}$ and $x_{2}$
are the coordinates of both constituents. We deal with two $8-$
spaces in the model: $(x_{1},x_{2})$ and $(x,y),$ where $x$ 
belonging to the Minkowski space $R^{4}_{1,3}$ are coordinates of 
a centre of inertia, and $y$ are internal coordinates. 
The author has assumed in \cite{2} that $y$ are transformed 
independently $x.$
\par
The equations of motion of such the composite system have the 
following algebraic property: they permit eight different solutions
of the kind $G_{A}\psi(x,y)$ if $\psi(x,y)$ is
some solution. Matrices $G_{A},\ A=1,...,8,$
set up a representation of the discrete group $D_{4}.$
If one assumes that transitions between these solutions are induced 
by transformations of a space $(y),$ it leads that an algebraic 
structure of field equations puts hard restrictions on this space: 
the space $(y)$ should be discrete.
\par
There are two possibilities: to have $y \equiv 0$ for all $A$ or to 
have two isolated sets of solutions when $y \not\equiv 0.$ To get 
the global symmetry group $SU(3)_{c}\times SU(2)_{l}$ 
for the model in the latter case, one must introduce an additional 
postulate about conservation of the norm of a set of solutions (see
\cite{2}, section 7). The internal coordinates $y$ are not observable; 
namely it leads to the apparent continuous system's symmetry for any
observer in the $(x)$ space. Some other features of the standard model
on global level are reflected by this model automatically. A minimal 
set of generations contains $4$ generations. A multiplet of any
generation contains two $SU(2)-$singlets that gives a possibility to
introduce non-zero neutrino masses that is important after their 
observation by the Super-Kamiokande collaboration \cite{4}.

\section[3]{Symmetries of the model of gravity in a flat $12-$space 
\cite{3}}
To embed the general relativity curved $4-$space into
a flat  $12-$space $(x,A,B)$ with flat $4-$sectors $(x), (A),
(B),$ one may map $(x) \rightarrow (A) \rightarrow (B) $
and get the connection $\Gamma = \tilde \Gamma + \bar \Gamma$  
with $\tilde \Gamma$ being its tensor part. This connection has a 
trivial curvature. After it, we may linear deform the connection:
$\Gamma \rightarrow \gamma = f\tilde \Gamma + \bar \Gamma $ where $f$ 
is a scalar parameter. A non-trivial curvature of the 
connection $\gamma$ is proportional to $(f^{2}-f).$ The last step is
to oblige the according Ricci tensor to Einstein's equations to get 
an alternative model of gravitation in the flat $12-$space \cite{3}. 
\par
In this case, $f$ is $U(1)-$symmetry's parameter; the parameter 
$F \equiv f^{2}-f$ gives a curvature scale. 
Two-valueness of the mapping $F \rightarrow f$ gives $D_{1}-$symmetry 
which may be transformed into the $SU(2)-$one as in the previous case. 
But the parameters of the $U(1)-$ and $SU(2)-$transformations will depend 
from each other in a general case. We have here a very interesting 
analogy with the standard model: to provide its independence, one should 
do a rotation on some angle $\theta$ in the parameter plane $(f,F)$ 
\cite{3}. If we take in the mind a necessity to unify the two described
models then the minimum value $\theta_{min}$ is determined by the demand 
that one component of the $SU(2)-$doublet should stay massless or near 
to it. We will have in such the case: $\sin^{2}\theta_{min} = 0,20$ 
from the pure geometrical reasons. $\theta_{min} $ is an analog of the 
Weinberg angle $\theta_{w}$ for which $\sin^{2}\theta_{w} = 0,215$ 
from an experiment. It is exiting that these values of angles 
$\theta_{min}$ and $\theta_{w}$ are approximate enough. After the 
rotation, the $SU(2)-$symmetry will be broken if we take
gravitation into account. I would like to note here that as it was
shown in my paper \cite{5}, quantum gravity would be super-strong on 
small distances of the order $\sim 10^{-11} \ m.$ It would mean that one
cannot consider it alone on this scale.

\section[4]{Conclusion } 

In the two cases, which are described here, one can use some constructive
(algebraic in both situations) features of the model yielding discrete
symmetries to get some continuous ones. Today, these two models are not 
connected between themselves. I think that to do the first step to 
unify them, one should linearize Eqs. (9) or (11) from \cite{3} which are
uniform relative to the connection $\tilde \Gamma.$ As the second step, I
consider now a possibility to assume that super-strong interacting 
gravitons \cite{5} are constituents of the composite fermions.


\begin{thebibliography}{References                        }
\bibitem{1}
Cheng, T.-P., Li, L.-F. Gauge theory of elementary particle
physics. Oxford: Clarendon Press, 1984.
\bibitem{2}
M.A.Ivanov. Nuovo Cimento, {\bf 105A} (1992) 77 [hep-th/0207210].
\bibitem{3}
M.A.Ivanov, General Relativity and Gravitation, {\bf 31} (1999) 1431
[gr-qc/0207017].
\bibitem{4}
Y.Fukuda et al. Phys. Rev. Lett., {\bf 81} (1998) 1562.
\bibitem{5}
M.A.Ivanov. Screening the graviton background, graviton pairing, and
Newtonian gravity [gr-qc/0207006].

\end{thebibliography}
\end{document}